\documentclass[3p,twocolumn,12pt]{elsarticle}

\usepackage{amssymb}
\usepackage{multicol}

\journal{Chem. Phys. Lett.}

\begin{document}

\begin{frontmatter}



\title{A quantum informational approach for dissecting chemical reactions}


\author[McMaster]{Corinne Duperrouzel}
\author[McMaster]{Pawe{\l} Tecmer\corref{cor1}}
\cortext[cor1]{(905) 525-9140, ext. 24162}
\ead{tecmer@mcmaster.ca}
\author[McMaster]{Katharina Boguslawski\corref{cor1}}
\ead{bogusl@mcmaster.ca}
\author[Landulet]{Gergerly Barcza}
\author[Landulet]{\"Ors Legeza}
\author[McMaster]{Paul W. Ayers}
\address[McMaster]{Department of Chemistry and Chemical Biology, McMaster University, Hamilton, 1280 Main Street West, L8S 4M1, Canada}
\address[Landulet]{Strongly Correlated Systems ``Lend\"ulet" Research Group, Wigner Research Center for Physics, H-1525 Budapest, Hungary}

\begin{abstract}
We present a conceptionally different approach to dissect bond-formation processes in metal-driven catalysis using concepts from quantum information theory. Our method uses the entanglement and correlation among molecular orbitals to analyze changes in electronic structure that accompany chemical processes. As a proof-of-principle example, the evolution of nickel-ethene bond-formation is dissected which allows us to monitor the interplay of back-bonding and $\pi$-donation along the reaction coordinate. Furthermore, the reaction pathway of nickel-ethene complexation is analyzed using quantum chemistry methods revealing the presence of a transition state. Our study supports the crucial role of metal-to-ligand back-donation in the bond-forming process of nickel-ethene.

\end{abstract}


\end{frontmatter}

\section{Introduction} 
\begin{figure}[b]
\centering
\includegraphics[width=0.75\linewidth]{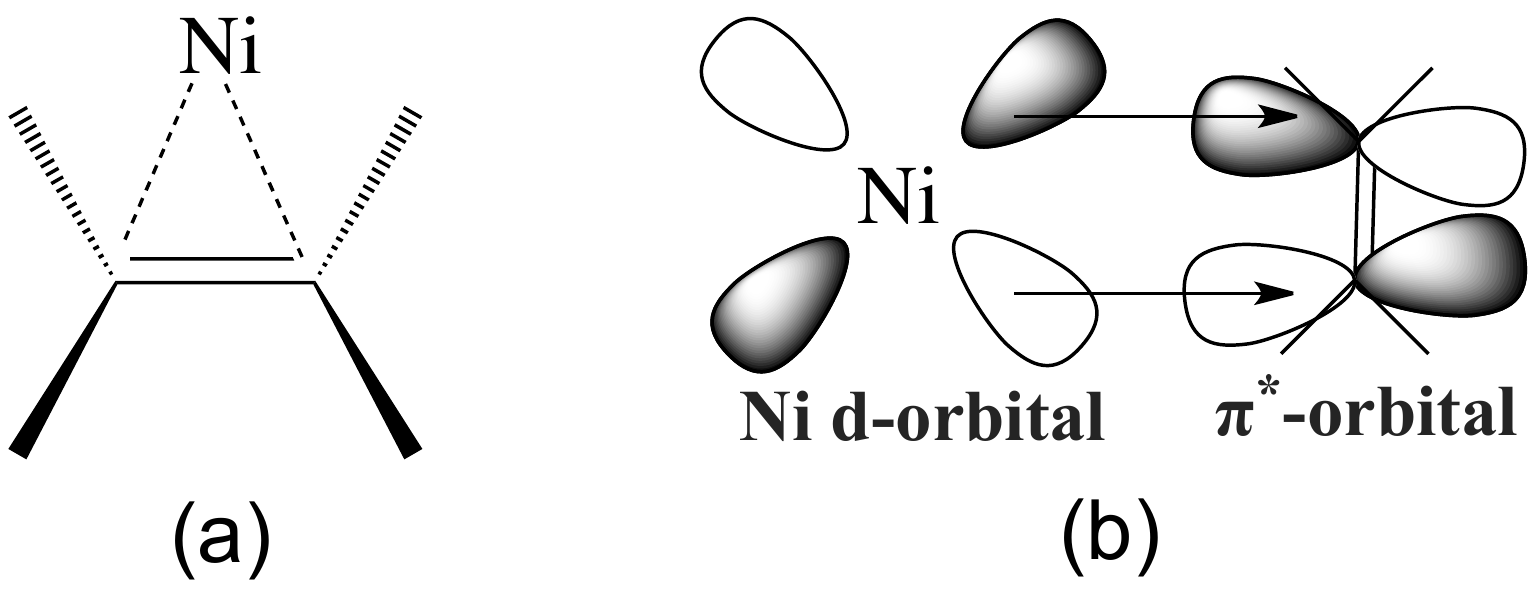}
\caption{(a) Lewis structures of nickel-ethene. (b) Schematic representation of back-donation.
}
\label{fig:structures}
\end{figure}

Nickel-ethene complexes have long been of interest in the field of organometallic chemistry. They have been used as localized models to study the chemisorption of ethene onto the surface of transition metals~\cite{McKee,RoschHoffmann1974,Ozin1978,Widmark1985}. The formation of a metal--olefin bond is made possible through the process of back-donation, in which nickel $d$-orbitals push electrons into the $\pi^*$-orbitals of ethene (see Figure~\ref{fig:structures}(b)). This transfer of electrons destabilizes the carbon-carbon double bond, allowing the complex to be used as a homogeneous catalyst for thermally forbidden processes, specifically [$2+2$] cycloaddition reactions which yield cyclobutane derivatives\cite{Mitsudo1979,Mitsudo1994}. 

The back-bonding phenomenon is commonly understood using molecular orbital diagrams, where filled metal $d$-orbitals interact with empty ligand $\pi^*$-orbitals. This donation of electrons from metal to ligand causes a flow of electron density towards the ligands. Molecular orbital theory, however, restricts the understanding of back-donation to a rather simplified and solely qualitative picture, hampering a detailed analysis of metal--ligand bond-formation processes and the specific role of back-bonding therein. 
To obtain a trustworthy qualitative picture, one first must perform accurate computational studies, and then post-process the complicated information encoded in the correlated wavefunction in a way that facilitates chemical interpretation.  
Popular analysis tools in quantum chemistry are, for instance, population analysis, spin and electron density distributions, and local spins.
We have recently developed a new approach, where concepts from quantum information theory are used to provide insight into molecular electronic structure and the changes in electronic structure that accompany chemical processes. Specifically, the quantum entanglement of, and between, orbitals can be used to extract bond orders~\cite{entanglement_bonding_2013,PCCP_bonding}, dissect weak interactions~\cite{CUO_DMRG}, and analyze orbital interactions~\cite{entanglement_letter,kurashige2013}. Importantly, this approach can be reliably applied whenever the electronic wavefunction can be accurately determined, even when the simple picture of interacting orbitals completely fails.
 
\begin{figure}[htp!]
\centering
\includegraphics[width=0.9\linewidth]{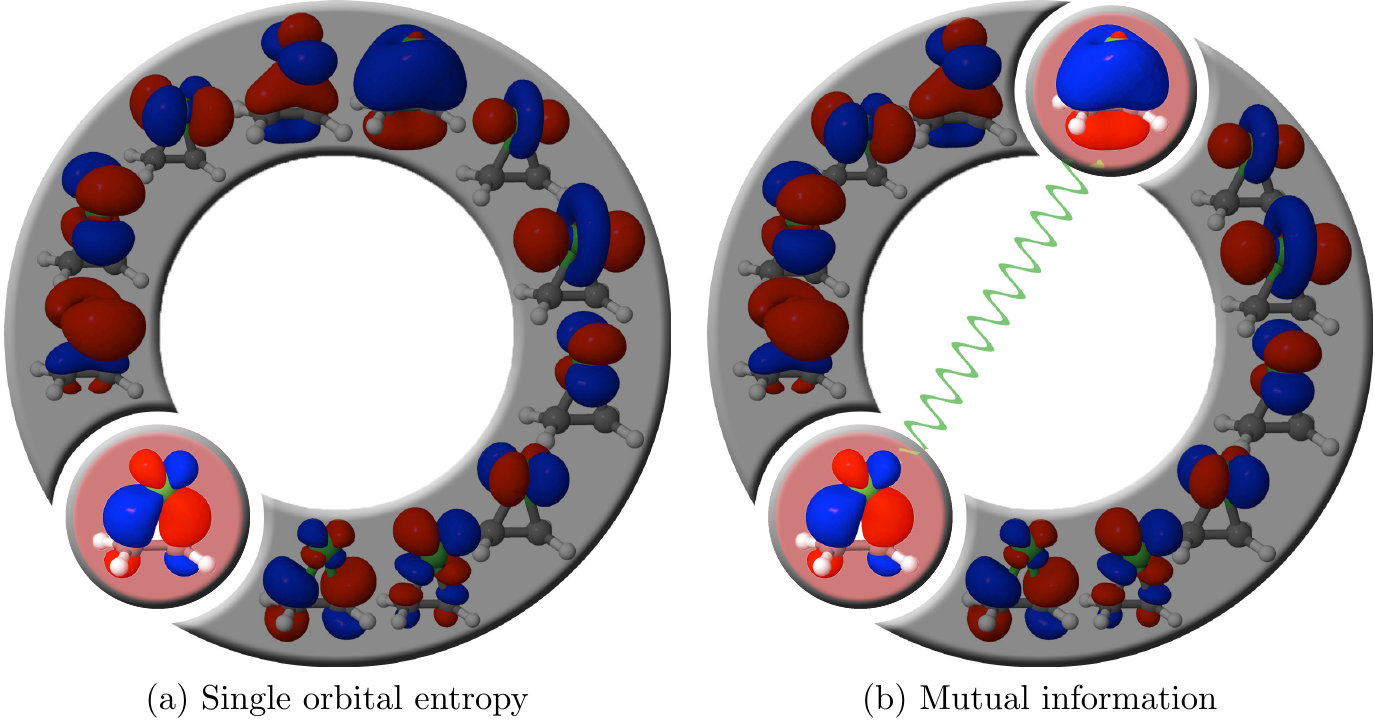}
\caption{Schematic representation of orbital entanglement. The orbitals in question are highlighted in red. The single orbital entropy and mutual information are evaluated using the von Neumann entropy. The mutual information is indicated as the green line in (b). 
}
\label{fig:entanglement}
\end{figure}

The entanglement between molecular orbitals is quantified using two entropic measures: the single orbital entropy and orbital-pair mutual information~\cite{legeza_dbss,Rissler,barcza2014entanglement}. Both entanglement measures quantify the correlation of individual orbitals or pairs of orbitals embedded in an orbital space (for instance, the active space in \emph{ab initio} calculations). Orbital entanglement can thus be considered as an alternative to molecular orbital theory, providing quantitative assessment of orbital interactions. A schematic representation of the single orbital entropy and mutual information is shown in Figure~\ref{fig:entanglement}. Specifically, the single orbital entropy measures how strongly an orbital (encircled in Figure~\ref{fig:entanglement}(a)) is entangled with the remaining orbitals (layered in grey in Figure~\ref{fig:entanglement}(a)). The single orbital entropy is large for orbitals with unpaired electrons (free radicals, open-shell states) and orbitals participating in chemical bonding~\cite{entanglement_bonding_2013,PCCP_bonding}. The mutual information quantifies the communication between orbital pairs (indicated by the green line in Figure~\ref{fig:entanglement}(b)). These correlations include both classical and quantum origin. Examples for strongly correlated orbital pairs are $\pi$- and $\pi^*$-orbitals and metal $d$-orbitals. Specifically, if two reactants approach each other, molecular orbitals that are involved in the bond formation process become strongly entangled. An orbital entanglement analysis is, thus, ideal to study molecular reactions because it allows us to identify and monitor the most important orbital interactions along the reaction coordinate. By analyzing the mutual information and single orbital entropy diagrams, we are able to decide which orbital interactions form first and at which points of the dissociation pathway orbitals start communicating with each other.

As transition metal compounds possess a non-negligible multireference character, they are a remarkable challenge for theoretical investigations. Several Complete-Active-Space Self-Consistent-Field (CASSCF) and Complete-Active-Space Self-Consistent-Field including second-order Perturbation Theory (CASPT2) studies have been conducted on metal-olefin complexes in the past\cite{Widmark1985, Siegbahn1986, Widmark1986, Widmark1987, Bernardi1997}, however the small active spaces used might not be ideal for an accurate description of transition metals. These limitations are alleviated by the Density Matrix Renormalization Group~\cite{white,PhysRevB.48.10345} (DMRG) algorithm that allows us to efficiently extend computational studies to large active spaces. The first application of DMRG to transition metal chemistry was presented by Marti \emph{et al.}\cite{marti2008}, demonstrating its superiority against conventional multireference approaches (see, for instance, Refs.~\citenum{marti2010b,ors_springer,chanreview,orbitalordering,fenoDMRG, Wouters2014, entanglement_letter} for recent examples).

In this paper, we investigate the reaction pathway of nickel-ethene complexation (see Figure~\ref{fig:structures}(a)) using quantum chemistry methods and demonstrate that the bond-formation process of nickel-ethene can be resolved by means of orbital entanglement. 
Specifically, orbital entanglement will allow us to determine exactly when back-donation starts to occur, and at which point of the reaction coordinate the bond between nickel and ethene forms. 

\section{Computational Details} 
\subsection{Geometry Optimization}
The structures of the nickel-ethylene were optimized using the \textsc{ADF} software package~\cite{adf1,adf2,adf2013}. We performed a constrained geometry optimization, where the bond lengths of the Ni--C dissociating centers were systematically varied while the structure of the ethylene subsystem was relaxed. The Ni--C distances were varied from 1.75\AA~ to 2.75\AA.  
In all calculations, the TZ2P~\cite{adf_b} basis set was used, along with the BP86 exchange--correlation functional~\cite{Perdew86,Becke}.
\subsection{CASSCF}
All CASSCF~\cite{Roos_casscf,Werner_1985,Knowles_1985} calculations were performed in the~\textsc{Dalton2013} program package~\cite{Dalton2013}. A TZP ANO-RCC basis set was used with the following contractions: H:$(8s4p3d1f)$~\cite{TZPhydrogen}, C:$(8s7p4d3f2g)\rightarrow[4s3p2d1f]$~\cite{TZPcarbon}, Ni:$(10s9p8d6f4g2h)\rightarrow[6s5p3d2f1g]$~\cite{TZPnickel}. 
Scalar relativistic effects were included through the second-order Douglas--Kroll--Hess Hamiltonian~\cite{Douglas_Kroll_1974,dkh2}.

The $2p_{\pi}$ and $2p_{\sigma}$ orbitals of ethylene, as well as the $3d$, $4s$ and $4d$ orbitals of the nickel atom were correlated, resulting in an active space of 12 electrons correlated in 12 orbitals (CAS(12,12)SCF). We have imposed C$_{2{ v}}$ point group symmetry for the nickel-ethylene complex.  The CAS(12,12)SCF orbitals are collected in Figure S1 in the Supporting Information~\cite{SI}.  
\subsection{DMRG}
The Budapest \textsc{DMRG} program~\cite{dmrg_ors} was used to perform the DMRG calculations. 
As orbital basis, the natural orbitals obtained from the CASSCF calculations as described in the previous subsection were used. 
The active spaces were extended by including additional occupied and virtual natural orbitals. 
Eight additional occupied orbitals (3 in $A_1$, 2 in $B_1$, 1 in $A_2$ and 2 in $B_2$) and sixteen virtual orbitals (5 in $A_1$, 5 in $B_1$, 2 in $A_2$ and 4 in $B_2$) were added to the active space, increasing it to 28 electrons correlated in 36 orbitals (DMRG(28,36)). 
 
To enhance convergence, we optimized the orbital ordering~\cite{orbitalordering} and applied the dynamic block state selection (DBSS) approach~\cite{legeza_dbss2,legeza_dbss3}; the initial guess was generated using the dynamically extended-active-space procedure (DEAS)~\cite{legeza_dbss}. 
In all DMRG calculations, the quantum information loss for the DBSS procedure was set to 10$^{-5}$ and the Davidson diagonalization threshold to 10$^{-6}$. The minimum number of block states, $m$, was set to 64, while the maximum number was set to 2048. The convergence of DMRG with respect to $m$ is summarized in Table~S1 of the Supporting Information~\cite{SI}. 

\section{Results and Discussion} 
\begin{figure*}[ht!]
\centering
\includegraphics[width=0.9\textwidth]{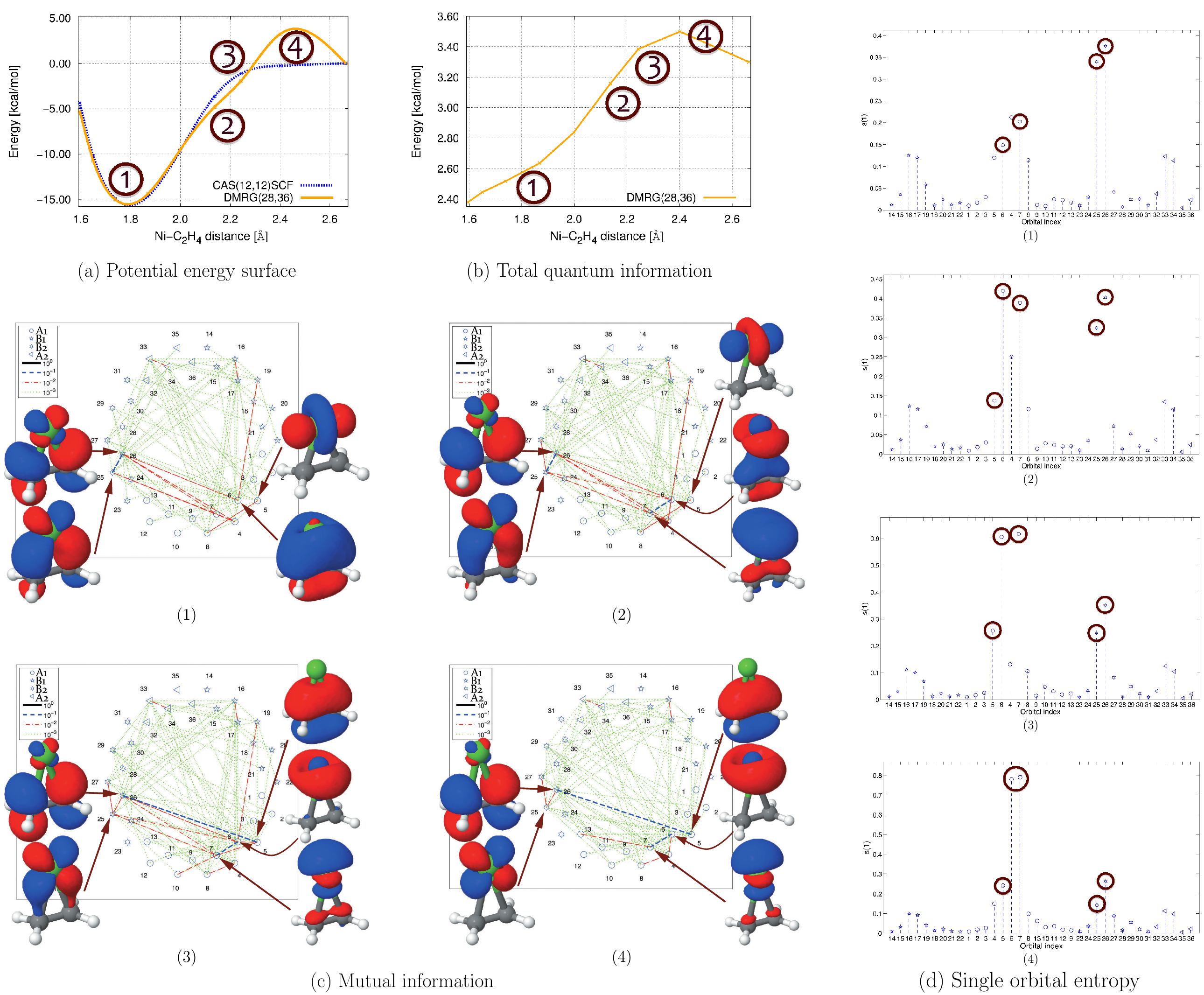}
\caption{(a) Potential energy surface and (b-d) quantum information analysis of nickel-ethene complexation with respect to selected nickel-ethene distances. The chosen distances (1-4) are marked in (a) and cover the transition state (1), and the equilibrium distance (4). Molecular orbitals are marked by different symbols (according to their irreducible representation) in the mutual information and single orbital entropy diagrams. Highly entangled orbitals are highlighted in (c) and (d).
}
\label{fig:3}
\end{figure*}
To react with ethene, the nickel atom has to undergo a spin-state change from a triplet (ground state of the free atom) to a singlet. The nickel-ethene reaction pathway describes, thus, an overall singlet state.
Potential energy surfaces determined from CASSCF and DMRG are shown in Figure~\ref{fig:3}(a). Both methods produce similar potential energy surfaces around the equilibrium distance. While the CASSCF potential energy surface is a purely dissociative curve, the DMRG potential energy surface predicts a transition state around a stretched nickel-ethene distance of 2.50~\AA{}. Specifically, CASSCF 
fails to locate the correct electronic state as the distance between the nickel atom and the carbon atoms increases. This qualitative error in the CASSCF wavefunction is reflected in the natural orbital occupation numbers. 
While CASSCF predicts occupation numbers of 1.30 and 0.70 for the nickel $d$-orbitals, DMRG yields occupation numbers of 1.00, \emph{i.e.}, two singly-occupied orbitals that are coupled to a singlet state on the nickel atom. Furthermore, an \emph{a posteriori} N-Electron-Valence 2nd-order Perturbation Theory (NEVPT2) correction of dynamic correlation does not cure the failure of CASSCF. We should note that DFT dramatically fails in predicting the dissociation pathway of nickel-ethene. 
Specifically, compared to multi-reference approaches, DFT overstabilizes and overbinds nickel-ethene by more than 35 kcal/mol and 0.1~\AA{}, respectively.
(The total failure of DFT methods for this reaction indicates that traditional arguments based on electron configurations and molecular orbitals are completely inappropriate for this reaction, see Figure S2.)

The presence of the transition state is also supported by quantum information theory. As emphasized in Ref.~\citenum{TTNS-LiF}, the behavior of the total quantum information---the sum of all single orbital entropies---as a function of bond distance allows us to detect and locate points where the electronic wavefunction changes drastically. Figure~\ref{fig:3}(b) shows the evolution of the total quantum information with respect to the nickel-ethene bond length. If the nickel atom and ethene molecule are pulled apart, the total quantum information increases gradually, up to a nickel-ethene distance of 2.4~\AA{}. Beyond this point, the total quantum information decreases indicating the transition state.

The bond-formation pathway and the presence of back-donation can be understood in terms of entangled molecular orbitals. The orbital entanglement diagrams for selected points of the dissociation pathway are collected in Figures~\ref{fig:3}(c) and \ref{fig:3}(d). The strength of the mutual information is color-coded: highly entangled orbitals are indicated by blue lines, while red and green lines indicate decreasing levels of entanglement. Furthermore, molecular orbitals that are involved in the bond formation process, including $\pi$-donation and back-donation, (ethene $\pi$- and $\pi^*$-orbitals and nickel $3d$/$4s$-orbitals) are highlighted in the mutual information diagram.

Close to the dissociation limit, the valence nickel and ethene orbitals are not entangled with each other (see Supporting Information~\cite{SI}) and the strongly entangled orbitals are the ethene $\pi$/$\pi^*$-orbitals (nos. 5 and 26) as well as the nickel $3d_{z^2}$/$4s$-orbitals (nos.~6 and 7). If the ethene molecule approaches the nickel atom (point (4) of the reaction coordinate), the nickel $3d_{xz}$-orbital becomes entangled with the ethene $\pi^*$-orbital (indicated by the red line between orbital nos.~25 and 26 in Figure~\ref{fig:3}(c-4)). Yet, the strongly entangled orbitals remain centered around the nickel atom and the ethene molecule (indicated by the blue lines on the mutual information diagram and the large values of the single orbital entropy). As the distance between nickel and ethene decreases and the system moves past the transition state (point (3) of the reaction coordinate), the nickel $3d$/$4s$-orbitals and the ethene $\pi$-orbital become strongly entangled (see Figure~\ref{fig:3}(c-3) and (d-3)) indicating the formation of the nickel-ethene bond through $\pi$-donation. Once the nickel-ethene distance decreases to approximately 2.15 \AA{} (point (2) of the reaction coordinate), we observe a transition of orbital entanglement where the pair of bonding and anti-bonding orbitals involved in back-donation (the nickel $3d_{xz}$ and ethene $\pi^*$-orbitals) as well as orbitals involved in $\pi$-donation (the nickel $3d_{z^2}$/$4s$- and ethene $\pi$-orbitals) become highly entangled. At this point, the orbital entanglement is dominated by the bonding interaction between the nickel atom and the ethene molecule, while, around the equilibrium distance, only orbitals involved in back-donation (nos.~25 and 26) remain strongly entangled. 
\section{Conclusions}
Our analysis demonstrates that concepts from quantum information theory constitute a promising tool to investigate well-established concepts of chemistry. Orbital entanglement allows us to monitor bond-formation processes and to highlight the most important orbital interactions along the reaction pathway. In the case of nickel-ethene complexation, metal-ligand bonding is initialized by back-donation which establishes around the transition state. This back-bonding then entails $\pi$-donation from the ethene ligand to the metal center. In addition, our study emphasizes the failures of routinely used DFT, CASSCF, and NEVPT2 approaches to describe the chemistry of compounds involving open-shell transition metals.

\textbf{Acknowledgments}. We gratefully acknowledge financial support from the Natural Sciences and Engineering Research Council of Canada and the Hungarian Research Fund (OTKA K100908 and NN110360). C.D.~acknowledge financial support from the McMaster Chemistry \& Chemical Biology Summer Research Scholarship.
K.B.~acknowledges the financial support from the Swiss National Science Foundation (P2EZP2 148650). 
\"O.L. acknowledges support from the Alexander von Humboldt foundation and from ETH Z\"urich during his time as a visiting professor.
The authors acknowledge support for computational resources from \textsc{SHARCNET}.


\bibliographystyle{achemso} 
\bibliography{rsc}

\end{document}